# Frequency manipulation of light by photonic gauge potentials

Chengzhi Qin[†,1,2], Feng Zhou[†,2], Yugui Peng[†,1,3], Hao Chen[1,2], Xuefeng Zhu[1,3], Bing Wang[1,2*], Jianji Dong[2*], Xinliang Zhang[2] and Peixiang Lu[1,2,4*]

[1]School of Physics, Huazhong University of Science and Technology, Wuhan, 430074, China

[2]Wuhan National Laboratory for Optoelectronics, Huazhong University of Science and Technology, Wuhan, 430074, China

[3]Innovation Institute, Huazhong University of Science and Technology, Wuhan, 430074, China

[4]Laboratory for Optical Information Technology, Wuhan Institute of Technology, Wuhan, 430205, China

[*]e-mails: wangbing@hust.edu.cn    jjdong@hust.edu.cn    and lupeixiang@hust.edu.cn
[†]These authors contributed equally to this work.

**The ability to manipulate the frequency of light is of great importance in both fundamental quantum sciences[1-3] and practical applications[4-9]. Traditional method for frequency conversion relies on nonlinear optical processes, which are faced with the obstacles of low efficiency and limited bandwidth[10, 11]. Recent developments of topological photonics introduce the concepts of gauge potentials and magnetic fields to the realm of photons[12-16]. Here, we demonstrate versatile frequency manipulation of light via photonic gauge potentials in a fiber-optic communication system. The gauge potential of frequency dimension is realized by controlling the initial phase of electro-optic phase modulation. A maximum 50 GHz frequency shift and three-fold bandwidth expansion for frequency combs are achieved by choosing different gauge potentials. By adopting two cascaded phase modulators with different gauge potentials, we also realize "negative refraction" for frequency combs and frequency "perfect imaging" for arbitrarily input spectra. These results may pave the way towards versatile frequency management in quantum optics and classical optical communications.**

Frequency is a fundamental degree of freedom for light since it is basically related to the energy of a photon. The manipulation of frequency, such as frequency conversion (shift), spectral expansion and compression, has enabled a wide range of applications from high-speed communication[6,7,17] to precise metrology and spectroscopy[4, 18, 19]. For instance, frequency conversion is widely used both in quantum information processing[1-3] and classical optical communications[6,7,17]. Spectral expansion and compression techniques can also have potential applications in optical frequency comb generation, multiple-wavelength sources[5-7], and spectral-temporal coupling systems[8,9]. Traditional methods for frequency conversion rely on nonlinear optical processes[10,11], but the conversion efficiency is fairly low due to the inherent weak nonlinear susceptibilities in dielectrics. To realize efficient frequency shift, recent efforts have been paid to the methods of dynamic modulation, such as acousto-optic[20], electro-optic modulation[21] and mechanical vibrations[1]. To control the bandwidth of light, the concept of time lens with electro-optic modulation has also been adopted to realize spectral expansion and compression[22, 23]. While the time lens needs to work in cooperation with accurately controlled dispersion components, which is accompanied by the increase of system complexity. Hence, a simple and universal scheme to realize versatile frequency manipulation is required.

In this Letter, we utilize the concept of photonic gauge potential to manipulate the frequency of light. As firstly put forward in real space, photonic gauge potential can be used to control the propagation of light and realize topologically protected edge states[13-16]. It is then generalized from spatial to frequency dimension based on a dynamic modulated coupled-resonator array[24,25]. However, it is difficult to demonstrate in experiments due to the complicated on-chip fabrication techniques for dynamic resonator arrays. Here, we demonstrate photonic gauge potentials based on optical phase modulators (PMs) in a fiber-optic communication system, which is convenient and offers all components on hand. The initial phase of electro-optic modulation in the PM corresponds to an effective gauge potential in the frequency dimension. By proper choices of gauge potentials, we realize directional shift, expansion and "negative refraction" for frequency combs. Additionally, the gauge potentials can also be used to control the frequency comb generation and realize "frequency perfect imaging" for an arbitrarily input spectrum.

The photonic gauge potential can be implemented in a travelling-wave optical phase modulator constituted by $LiNbO_3$ waveguide. As shown schematically in Fig. 1a, the PM is driven by a sinusoidal radiofrequency (RF) signal with a frequency of $\Omega$. The refractive index of the waveguide is thus $n(z, t) = n_0 + \Delta n \cdot \cos(\Omega t - qz + \phi)$. Here $n_0$ is the background index with $\Delta n$ and $\phi$ being the amplitude and initial phase of index modulation. The length of the PM is denoted by $L$ and the phase modulation depth reads $m_\varphi = \Delta n k_0 L$ with $k_0$ being the wavenumber in air. The modulation depth is related with the voltage V of the RF signal by $m_\varphi = \pi V/V_\pi$ with $V_\pi$ being the half-wave voltage[26]. The wavenumber of the RF travelling wave complies with $q = n_0\Omega/c$ such that the phase-matching condition is automatically satisfied. Under the modulation, the intraband transition between adjacent modes with a frequency interval of $\Omega$ will occur. $a_n(z)$ represents the amplitude of $n$th-order mode with frequency $\omega_n = \omega_0 + n\Omega$ and wavenumber $\beta_n = \beta_0 + nq$ ($n = 0, \pm1, \pm2,\ldots$), which is governed by the coupled-mode equation (Supplementary Sec I)

$$i\frac{\partial a_n(z)}{\partial z} = C\left(e^{i\phi}a_{n-1}(z) + e^{-i\phi}a_{n+1}(z)\right), \tag{1}$$

where $C = \Delta n k_0/2$, denoting the coupling coefficient between adjacent order modes. The modulation depth is thus $m_\varphi = 2CL$. All discrete modes acquire an additional phase during the photonic transition, which equals to the modulation phase $\phi$ for the upward transitions and $-\phi$ for the downward (Fig.1b). Such a nonreciprocal phase is a photonic analogue of Peierls phase that corresponds to an effective gauge potential $A_{eff}$ governed by $\int_{\omega_n}^{\omega_{n+1}} A_{eff} d\omega = \phi$ [12-14]. Here the gauge potential is presented in the frequency dimension with a constant value $A_{eff} = \phi/\Omega$. Since the choice of $\phi$ relies on the time origin of modulation, the potential should be gauge-dependent and cannot be measured directly.

To explore the measurable effect of gauge potential in the frequency dimension, we consider a frequency comb with uniform comb lines incident into the PM. The amplitude of the frequency comb is denoted by $a_n(z) = a_0(0)\exp(in\phi_0)\exp(ik_z z)$, where $a_0(0)$ is the uniform amplitude and $\phi_0$ the phase difference between adjacent comb lines. $k_z$ is the collective propagation constant along $z$ direction. Note that the amplitude here is a Bloch mode in the frequency dimension with $\phi_0$ being the corresponding Bloch momentum. Substituting $a_n(z)$ into Eq. (1), we can obtain the band structure

$$k_z(\phi_0) = -2C\cos(\phi_0 - \phi), \tag{2}$$

The gauge potential can give rise to a shift of band structure by $\phi$ in momentum space of frequency (Fig. 1c). Consequently, the frequency comb will undergo a spectral shift (Supplementary Sec II)

$$\Delta\omega = -m_\varphi \Omega \sin(\phi_0 - \phi), \tag{3}$$

As the phase modulation depth is fixed in the PM, the total frequency shift is uniquely determined by the phase difference $\phi_0 - \phi$. For a given frequency comb with an initial Bloch momentum $\phi_0$, the gauge potential $\phi$ can thus control the spectral shift of the frequency comb.

We build up an optical-fiber circuit with two cascaded PMs to demonstrate the effect of gauge potentials. As shown schematically in Fig. 2a, the frequency comb is generated by a mode-locked laser. To facilitate the observation of frequency shift, the initial frequency comb is filtered by a tunable optical band-pass filter to form a Gaussian envelope. To ensure the mode-locked laser and PMs are phase-correlated, we synchronize them with the same RF signal at a fixed modulation frequency of $\Omega/2\pi = 10$ GHz. The phase difference between the mode-locked laser and PMs can be continuously adjusted by the RF phase shifter (PS0). The phase difference of modulation in the two PMs is varied by the RF phase shifters (PS1) and (PS2). (See Methods).

Firstly we keep the two PMs modulated in phase such that $\phi_1 = \phi_2$. As the total modulation depth $m_\varphi$ remains constant, the frequency comb experiences a sinusoidal variation when the phase difference $\phi_0 - \phi_1$ raises incrementally from 0 to $3\pi$ (Fig. 2b). The frequency shift is exactly due to the variation of gauge potential, which agrees fairly with the theoretical prediction in Eq. (3). We can also fix the phase difference and successively increase the modulation depth (See Methods). It shows that the comb envelope manifests linear blue and red shifts as $\phi_0 - \phi_1 = -\pi/2$ and $\pi/2$, respectively (Figs. 2c and 2d). The maximum frequency shift is ~ 50 GHz (~ 0.4 nm) which spans up to five comb lines. When the phase difference becomes $\phi_0 - \phi_1 = 0$, the center of the frequency comb does not change. However, the spectrum is broadened with the full width at half maximum (FWHM) of Gaussian envelope increasing by three folds from 0.18 to ~ 0.6 nm (Fig. 2e). The spectral broadening is determined by the diffraction coefficient $D = 2\Omega^2\cos(\phi_0 - \phi)$[27]. For $\phi_0 - \phi_1 = \pm \pi/2$, we have $D = 0$, the frequency comb thus experiences diffraction-free evolution and remains a constant spectral width. While for $\phi_0 - \phi_1 = 0$ or $\pi$, the diffraction coefficient reaches the maximum $D = 2\Omega^2$ and the spectrum has largest width. The process of frequency evolution here is similar to the discrete diffraction of light in the waveguide arrays[27, 28]. By applying distinct gauge potentials in the two PMs such that $\phi_1 \neq \phi_2$, we can realize different routings for frequency combs such as "negative" and "positive" refraction (Supplementary Sec II). Specifically for $\phi_0 - \phi_1 = -\pi/2$ and $\phi_0 - \phi_2 = \pi/2$ (Fig. 2f), the frequency comb undergoes a "negative refraction" at the connection of two PMs. The deviation of frequency shift can thus be completely compensated by such an out-of-phase modulation in the PMs.

Apart from controlling the evolution of frequency comb, the gauge potentials can also be used to control the frequency comb generation. For one PM under a single frequency input, the output spectrum is $a_n = (i)^n J_n(m_\varphi)\exp(in\phi)$. Since the definition of modulation phase $\phi$ depends on the choice of time origin, it has no physical meaning and can't be detected from the amplitude spectrum. Here we cascade two PMs to control the frequency comb generation by the phase difference of modulation. The experimental setup is shown in Fig. 3a where a continuous-wave laser with the wavelength of 1550 nm is incident into the PMs. Firstly we fix the modulation depth as $m_\varphi$ in each PM and vary the phase difference of modulation $\Delta\phi = \phi_2 - \phi_1$. The output spectrum is illustrated in Fig. 3b, which can be precisely described by $|a_n| = |J_n(2m_\varphi\cos(\Delta\phi/2))|$ (Supplementary Sec III). As $\Delta\phi$ varies from 0 to $\pi$, the width of generated frequency comb can be squeezed from the maximum to a single frequency.

We can also fix the phase difference of modulation and vary the modulation depths $m_{\varphi 1}$ and $m_{\varphi 2}$ individually. For in-phase modulation $\Delta\phi = 0$ (Fig. 3c), the output spectrum is $|a_n| = |J_n(m_{\varphi 1} + m_{\varphi 2})|$, which exhibits constructive interference of frequency comb generation in the two PMs. While for out-of-phase modulation $\Delta\phi = \pi$ (Fig. 3d), the output spectrum is $|a_n| = |J_n(m_{\varphi 1} - m_{\varphi 2})|$, which manifests destructive interference of comb generation in the PMs. Interestingly as $m_{\varphi 1} = m_{\varphi 2}$, the output spectrum restores to the input single frequency, exhibiting the effect of "frequency perfect imaging". The perfect imaging can be understood in terms of the band structure (Fig. 3e). Based on the concept of Fourier transformation, a single frequency can be decomposed into a series of frequency combs with Bloch momenta covering the whole Brillouin zone. Due to the up-down symmetry of band structures for $\Delta\phi = \pi$, every frequency comb with Bloch momentum $\phi_0$ will experience a "negative refraction" with opposite phase front evolutions in the two PMs (Supplementary Sec III). So the single input frequency can be perfectly recovered at the output. The frequency perfect imaging here is analogous to spatial perfect imaging by a "superlens" realized by negative refraction[29]. While differing from the real-space "superlens" which relies on negative refractive index material, the "frequency superlens" here can be flexibly constructed by out-of-phase modulation. Actually, the perfect imaging can be generalized to arbitrary input spectra, including both discrete and continuous ones. Here we use supercontinuum source as input, the frequency components of which are incoherent[30] (Fig. 3a). We firstly keep equal modulation depth in the PMs and vary the phase difference $\Delta\phi$ from 0 to $\pi$ (Fig. 4a), the output spectral width is squeezed from ~ 0.8 nm to 0.18 nm. For $\Delta\phi = 0$ (Fig. 4b), the supercontinuum spectrum experiences linear broadening as the modulation depth increases. While for $\Delta\phi = \pi$ (Fig. 4c), the spectrum is broadened in the first PM and compressed in the second, ultimately restoring to the initial profile.

Finally, we consider the nonreciprocal property of spectrum evolution in the system. Keeping the first PM unchanged, we revert light propagation in the second PM such that the RF and optical modes propagate in opposite directions (Fig. 5a). It shows the frequency comb generated in the first PM remains nearly unchanged in the second PM. The nonreciprocal property is attributed to the phase-matching condition for the photonic transitions. As the RF and optical waves propagate in opposite directions, phase-matching condition is destroyed and photonic transition is negligible. In Figs. 5b and 5c, we input single frequency into one PM in forward and backward directions and vary the modulation frequency from 6 GHz to 18 GHz. As light propagates in the forward, the PM can efficiently generate a frequency comb with the frequency interval equal to the modulation frequency. While in the backward, the side bands of output spectrum are negligible. These results indicate that the nonreciprocal property holds for a rarely broad bandwidth in the system.

In conclusion, we have built up an optical circuit with two cascaded PMs to realize photonic gauge potential in the frequency dimension, which is benefit to the frequency manipulation of light in telecommunication wavelength. The band structure of incident frequency comb can be shifted by the gauge potential and thus the frequency evolution can be artificially controlled. A maximum frequency shift up to 50 GHz and bandwidth expansion up to 3 folds for frequency combs are realized by choosing different gauge potentials. By choosing out-of-phase modulation in the two PMs, we also realize "negative refraction" for frequency combs and "perfect imaging" for arbitrary input spectra. The study opens a new window to manipulate the frequency of light and can enable the developments of high-efficient frequency shifters and "frequency superlens", which may have potentials in high-speed optical communication and signal processing.

**Methods**

**Experimental setups and measurements.** The RF components are: RF source (Agilent E8247C) with a bandwidth from 250 kHz to 20 GHz. Phase shifter (Weinschel 981) with a bandwidth from DC to 18 GHz and phase shifter range ~ 1350°. Step attenuator (narda 4745-69) with the step of 1dB from 0 to 69 dB and bandwidth from DC to 18 GHz. Electronic power amplifier (Mini-Circuits ZVE-3W-183+) with a typical power gain ~ 30 dB and bandwidth from 5.9 to 18 GHz. The optical components are: Phase modulator (EOSPACE PM-5V4-40-PFA-PFA-UV) with $V_\pi$ = 2.8 V (@1 GHz) and bandwidth up to 40 GHz. Continuous-wave laser (Alnair Labs TLG-200). Supercontinuum source (Amonics ALS-CL-15-B-FA). Optical filter (Finisar 1000S).

In the RF circuits, we utilize two sets of phase shifters and attenuators to balance the insertion losses for the two PMs since "frequency perfect imaging" requires $m_{\varphi 1} = m_{\varphi 2}$. The phase difference of modulation between the two PMs can be continuously tuned by one of the phase shifters PS1 or PS2. The individual phase modulation depth $m_{\varphi i} = \pi V_i/V_\pi$ ($i$ = 1, 2) in each PM can be varied by changing the driving voltage though the RF attenuator in each path. Specifically, we firstly turn off the second PM and increase the driving voltage for the first PM to a specific maximum value. Then we keep the first PM unchanged and increase the driving voltage in the second PM until the driving voltages are equal to each other.

In the optical circuits, the fiber polarization controllers (PCs) are used to tune the polarization state of the input light for the two PMs. EDFA is utilized to compensate the insertion loss of the PMs. Though the absolute optical power has no influence on the relative spectral profile, higher optical power can reduce the influence of noise.

**Acknowledgements**

The work is supported by the 973 Program (No. 2014CB921301), the National Natural Science Foundation of China (No. 11304108, 11674117, 61622502), Natural Science Foundation of Hubei Province (2015CFA040).

**Author contributions**

B. W., C. Z. Q. and P. X. L. conceived the idea. B. W., C. Z. Q., F. Z., Y. G. P., and H. C. designed and performed the experiment. C. Z. Q. and Y. G. P. analyzed the data. P. X. L., B. W. and J. J. D. supervised the project. All authors contributed to the discussion of the results and writing of the manuscript.

**Competing financial interests**

The authors declare no competing financial interests.

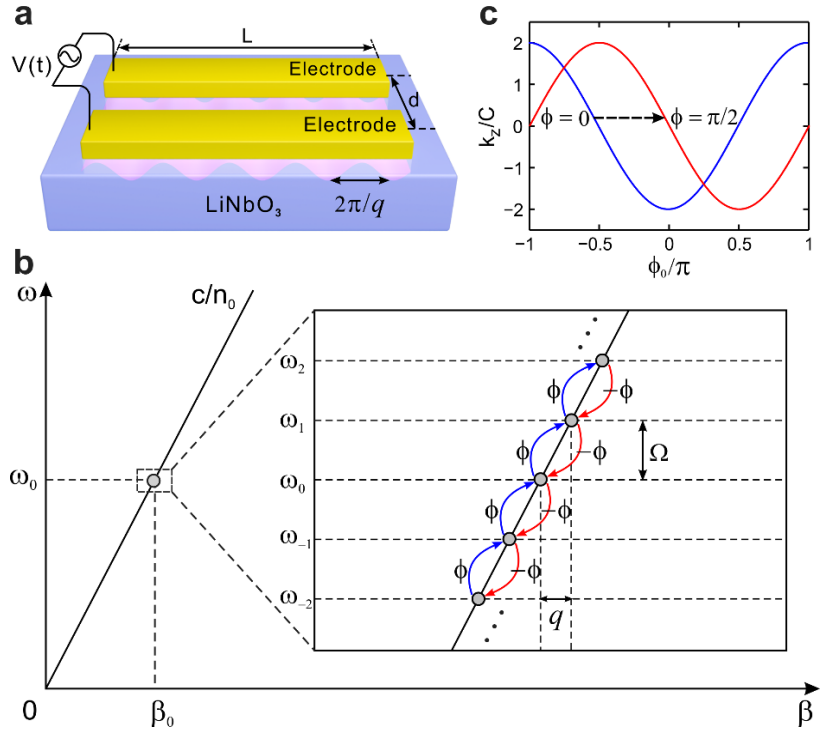

**Figure 1 | Photonic gauge potential in frequency dimension generated in phase modulator. a**, Schematic of a LiNbO$_3$ electro-optic phase modulator (PM). The refractive index of the substrate is modulated by a travelling-wave radiofrequency (RF) signal through the two parallel metal electrodes above. **b**, Photonic intraband transitions and effective gauge potential in the frequency dimension. The modulation creates a discrete frequency dimension with an interval of $\Omega$. The wavenumber of RF wave (denoted in pink) is $q = n_0\Omega/c$, which can compensate the wave vector mismatch in the photonic transitions. The optical mode acquires the modulation phase $\phi$ in the upward transition and $-\phi$ in downward, corresponding to an effective gauge potential $A_{eff} = \phi/\Omega$ in the frequency dimension. **c**, Band structure of frequency comb for $\phi = 0$ and $\phi = \pi/2$. The gauge potential can give rise to a shift of band structure in momentum space.

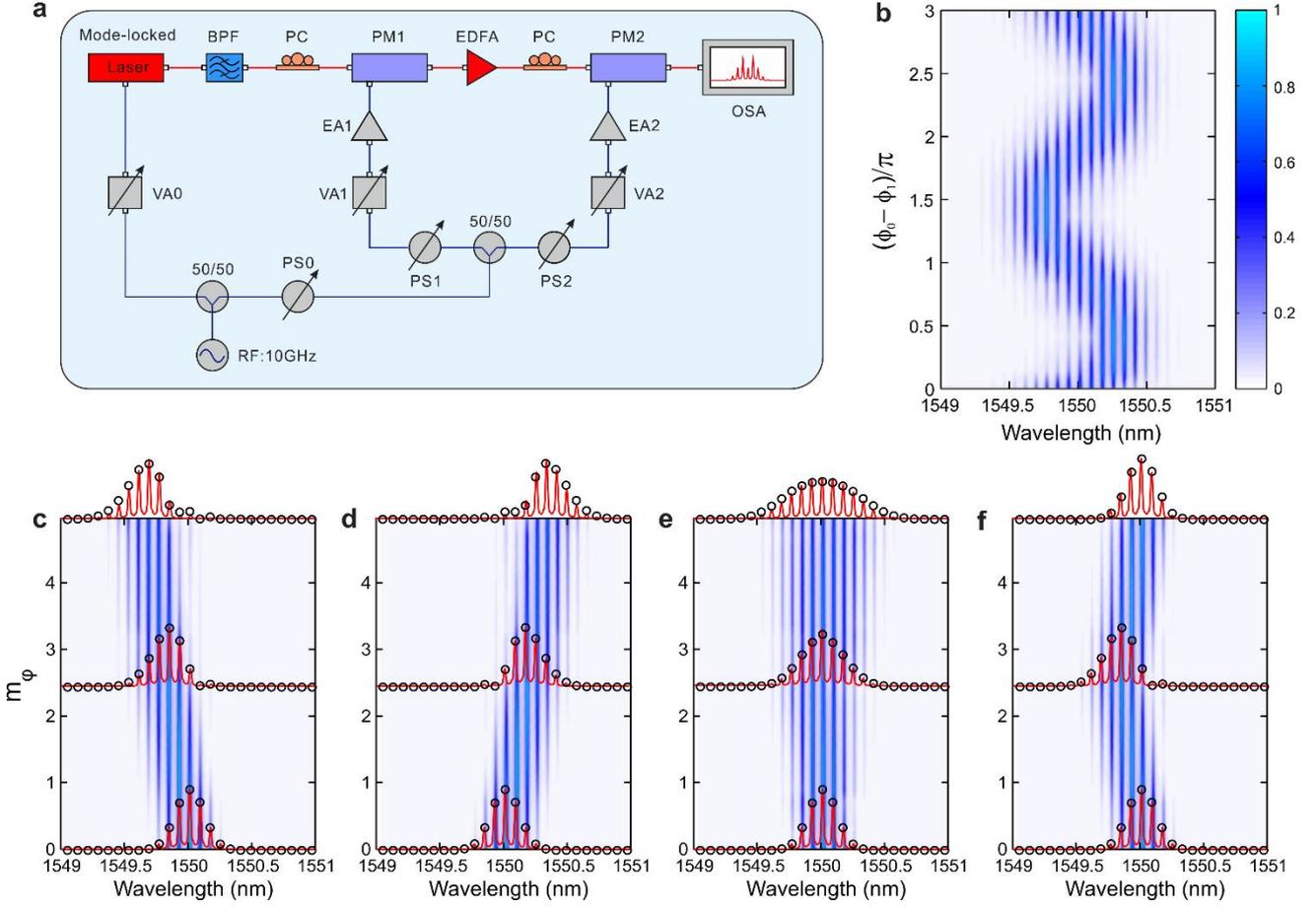

**Figure 2 | Frequency shift and bandwidth broadening by gauge potentials. a**, Schematic of the experimental set-up. RF, radiofrequency; 50/50, power splitter; PS, RF phase shifter; EA, electrical power amplifier; VA, variable RF attenuator; BPF, band pass filter; PC, polarizer controller; PM, phase modulator; EDFA, erbium-doped fiber amplifier; OSA, optical spectrum analyzer. The mode-locked laser and two PMs are driven by the same RF signal with a fixed frequency of $\Omega/2\pi =$ 10 GHz. **b**, Output spectral evolution as the phase difference $\phi_0 - \phi_1$ varies from 0 to $3\pi$. $\phi_0$ is the phase difference of adjacent orders in the mode-locked laser with $\phi_1$ being the phase of modulation in the two PMs. **c-e**, Output spectral evolution for fixed phase difference as the total modulation depths ($m_\varphi = m_{\varphi 1} + m_{\varphi 2}$) linearly increase. The experimental (red lines) and theoretical (black circles) frequency combs at the input, center and output of the two PMs are incorporated in the figures, which agree well with each other. The phase differences are $\phi_0 - \phi_1 = -\pi/2$, $\pi/2$ and 0 in **c**, **d** and **e**, corresponding to spectral blue shift, red shift and broadening, respectively. **f**, Frequency "negative refraction" as the two PMs are driven with different phases such that $\phi_0 - \phi_1 = -\pi/2$ and $\phi_0 - \phi_2 = \pi/2$.

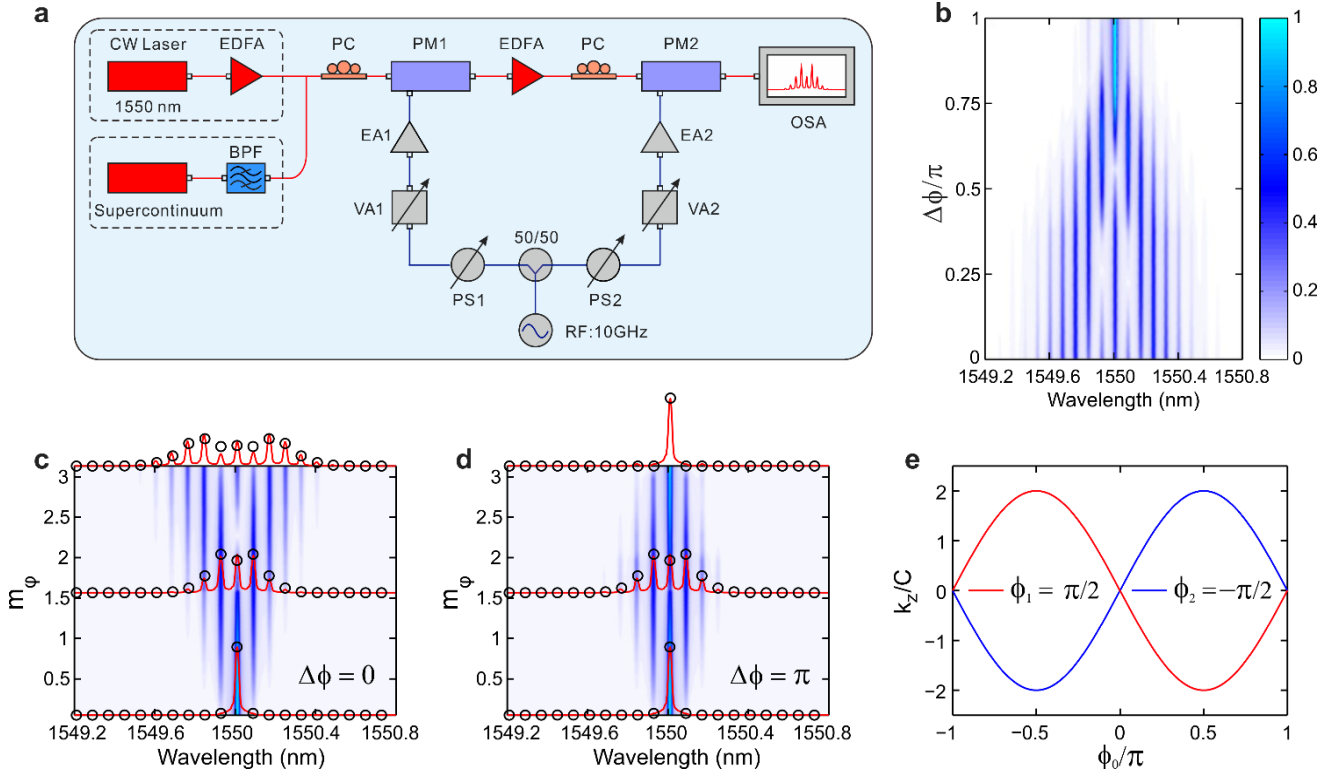

**Figure 3 | Frequency comb generation and frequency perfect imaging. a**, Schematic of the experimental set-up. CW laser, continuous-wave laser. The other components are the same with those in Fig. 2a. The CW laser and the supercontinuum source are input in the PMs, respectively. **b**, Output spectrum as a function of modulation phase difference $\Delta\phi$ in the two PMs. The spectrum is squeezed from widest to a single frequency as $\Delta\phi$ increases from 0 to $\pi$. **c**, Frequency comb generation for in-phase modulation $\Delta\phi = 0$ as the total phase modulation depth ($m_\varphi = m_{\varphi 1} + m_{\varphi 2}$) linearly increases. **d**, Frequency perfect imaging for out-of-phase modulation $\Delta\phi = \pi$. **e**, Band structures of frequency comb in the two PMs. Here we choose out-of-phase modulation $\phi_1 = -\pi/2$ and $\phi_2 = \pi/2$. For any frequency comb with Bloch momentum $\phi_0$, the propagation constants of comb amplitude in the two PMs satisfy $k_{z1}/k_{z2} = -1$, leading to frequency perfect imaging for a single frequency input.

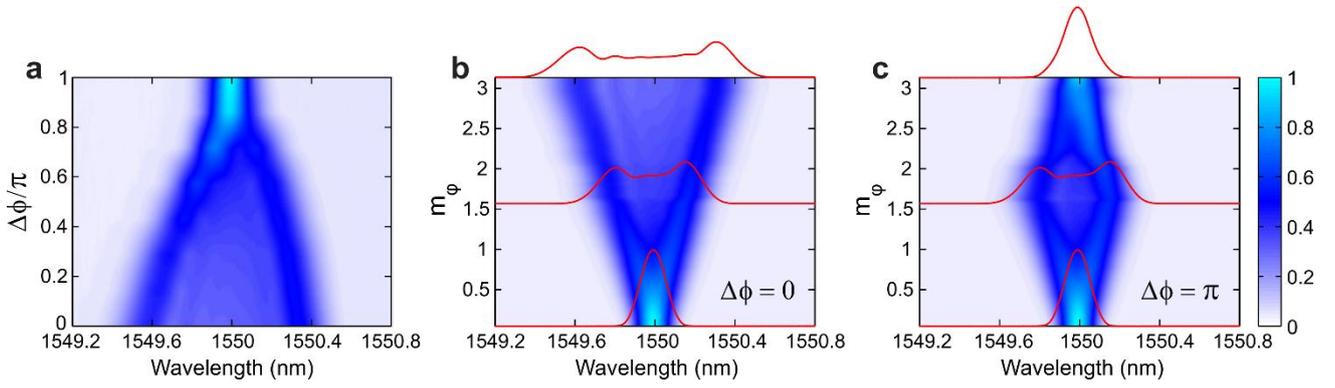

**Figure 4 | Frequency perfect imaging for supercontinuum spectrum. a,** Evolution of the output spectrum as the phase difference of modulation $\Delta\phi$ increases from 0 to $\pi$ for equal modulation depth in the two PMs. The input spectrum exhibits a Gaussion envelope with fixed FWHM of 0.18 nm. **b, c,** Output spectrum evolution under in-phase modulation $\Delta\phi = 0$ and out-of-phase modulation $\Delta\phi = \pi$, respectively. The red lines denote the spectra of input, at the center of two PMs and at the output end of the second PM, respectively.

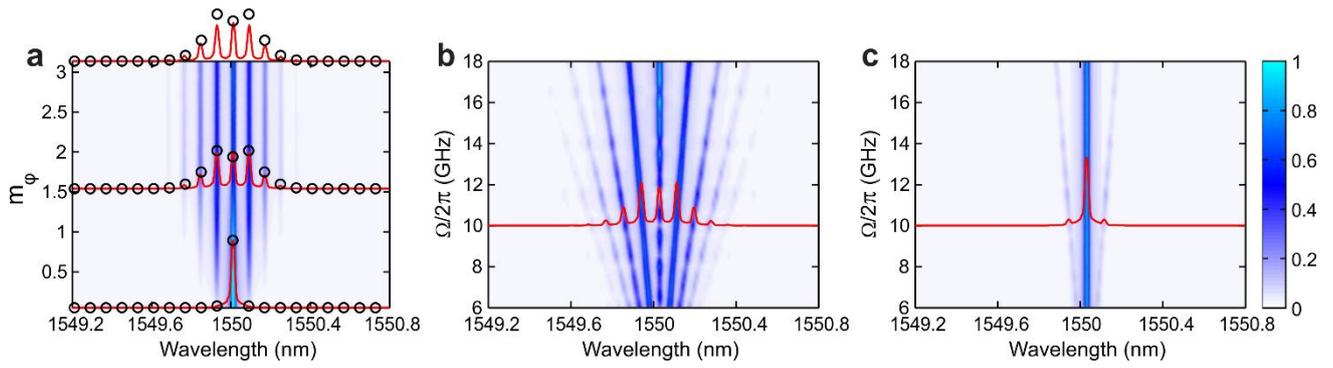

**Figure 5 | Nonreciprocal property of spectral evolution in the PM. a**, Output spectral evolution as light propagates in the forward direction in the first PM and backward direction in the second. The red lines and black circles denote the experimental and theoretical results, respectively. **b, c**, Output spectrum of one PM under the forward and backward input of a single frequency as the modulation frequency linearly increases from 6 GHz to 18 GHz. The red lines denote the case of $\Omega/2\pi = 10$ GHz, as demonstrated in the experiments. The PM can generate frequency comb only as the light is input in the forward direction, indicating the broadband nonreciprocal properties for the PM.